\documentclass[aps,preprint,showpacs,preprintnumbers,amsmath,amssymb,nofootinbib]{revtex4}

\usepackage{graphicx}
\usepackage{dcolumn}
\usepackage{bm}
\newcommand{\ybox}[2]   {
    \begin{center}
    \resizebox{!}{#1\textheight}
{\includegraphics{#2.eps}}
\end{center}           }
\def\refb#1{(\ref{#1})}

\begin{document}
\title{Uncertainties in limits on TeV-gravity from neutrino-induced showers}
\author{Eun-Joo Ahn}
    \altaffiliation{Email: sein@oddjob.uchicago.edu}
\affiliation{Department of Astronomy \& Astrophysics, 
and Kavli Institute for Cosmological Physics, University
of Chicago, 5640 S. Ellis Avenue, Chicago, IL 60637, USA}
\author{Marco Cavagli\`a}
    \altaffiliation{Email: cavaglia@phy.olemiss.edu}
\affiliation{Department of Physics and Astronomy, The University of
Mississippi, PO Box 1848, University, MS 38677-1848, USA}
\author{Angela V.\ Olinto}
    \altaffiliation{Email: olinto@oddjob.uchicago.edu}
\affiliation{Department of Astronomy \& Astrophysics, Enrico Fermi
Institute, and Kavli Institute for Cosmological Physics, University
of Chicago, 5640 S.\ Ellis Avenue, Chicago, IL 60637, USA}
\date{\today}
\begin{abstract}
In models with TeV-scale gravity, ultrahigh energy cosmic rays can generate
microscopic black holes in the collision with atmospheric and terrestrial
nuclei. It has been proposed that stringent bounds on TeV-scale gravity can be
obtained from the absence of neutrino cosmic ray showers mediated by black
holes. However, uncertainties in the cross section of black hole formation and,
most importantly, large uncertainties in the neutrino flux affects these
bounds. As long as the cosmic neutrino flux remains unknown, the
non-observation of neutrino induced showers implies less stringent limits
than present collider limits.
\end{abstract}
\maketitle
\section{Introduction}
It has been proposed that in models of TeV-scale gravity \cite{tev}, ultrahigh
energy cosmic neutrinos colliding with atmospheric nuclei could form black
holes (BHs) in the atmosphere \cite{Feng:2001ib}. (For a review on
nonperturbative gravitational events at trans-Planckian energies and
references, see e.g. Ref.~\cite{Cavaglia:2002si}.) The products of BH decay
would then be detected as extensive air showers (EASs) of ultrahigh energy
cosmic rays (UHECRs).

One of the signatures of BH formation in the atmosphere \cite{atmosphere,
Ahn:2003qn} would be the observation of deeply penetrating quasi-horizontal
EASs. Semiclassical calculations of BH cross sections suggest an interaction
length for neutrino-nucleon events in air of the order of $10^5$ g cm$^{-2}$,
about two orders of magnitude lower than the standard model (SM). This is
considered enough to generate deeply penetrating horizontal air showers with
little background from the SM. The nonobservation of these deeply penetrating
horizontal air showers can set an upper bound on the BH cross section
\cite{Anchordoqui:2001cg, Anchordoqui:2002vb, Anchordoqui:2003jr}. Since the
semiclassical BH cross section is inversely proportional to the fundamental
gravitational scale $M_D$, a lower bound on $M_D$ follows. For a number of
extra dimensions $n\ge 5$, the estimated lower bound on the gravitational scale
is $M_D > 1.0 - 1.4$ TeV \cite{Anchordoqui:2003jr}.

Limits on the fundamental gravitational scale from air showers depend on two
main assumptions: i) the existence of a cosmogenic  neutrino flux; and  ii) the
accuracy of estimates for the cross section for BH formation. The cosmogenic
neutrino flux is not known with good precision. Many models have been proposed
to estimate the flux, which varies by more than an order of magnitude (see e.g.
Refs.~\cite{cosmoflux, Protheroe:1995ft, Engel:2001hd}). Flux constraints made
from experiments are quite generous in their range \cite{Fodor:2003ph}. In
addition, microscopic BH formation at trans-Planckian energies \cite{blackhole}
is not understood nor has it been observed in particle collisions. Recent work
assumes that the cross section for BH formation at parton level is
approximately the semiclassical black disk (BD) area with Schwarzschild radius
$r_s(M)$, where $M$ is the center of mass (c.m.) energy of the collision.
However, a rigorous calculation of the BH cross section is still an open
question.

The detection of microscopic BH formation in particle collisions would confirm
TeV-gravity models. However, nonobservation of these events does not
necessarily rule out TeV-gravity at this stage. A more stringent constraint on
$M_D$ cannot be set until the neutrino flux and the physical process of BH
formation are better understood. Microscopic BHs might not form in the
atmosphere even if the gravitational scale is of order of the TeV. Therefore,
constraints on $M_D$ from nonobservation of atmospheric BH events are not as
stringent as what was previously considered. The perturbative predictions, on
which current collider bounds are based \cite{collider, Giudice:2003tu}, seem
to be a much stronger basis than those invoking strong gravitational effects.

In this article we revisit and discuss lower bounds on $M_D$ from
nonobservation of BH-induced EASs in more detail, and find how various
uncertainties may reduce these bounds.
\section{Cross section}
The total cross section for a neutrino-nucleon BH event in $(n+4)$-dimensions
is obtained by summing the neutrino-parton cross section of BH formation
$\sigma_{\nu i \rightarrow BH}(xs,n,M_D)$ over the parton distribution
functions (PDFs) $q_i(x,-Q^2)$ \cite{Brock:1993sz}:
\begin{equation}
\sigma_{\nu p \to BH}(s,x_m,n,M_D) = \sum_{i} \int_{x_m}^{1} dx \, q_i(x,-Q^2)
\, \sigma_{\nu i \rightarrow BH}(xs,n,M_D) \,,
\label{totcross}
\end{equation}
where $-Q^2$ is the four-momentum transfer squared, $\sqrt{x}$ is the fraction
of the nucleon momentum carried by the parton, $\sqrt{sx_m}=M_{BH,min}$, the
minimal BH mass where semiclassical description is valid, and the parton cross
section $\sigma_{\nu i \rightarrow BH}(xs,n,M_D)$ is given by
\begin{equation}
\displaystyle
\sigma_{\nu i \rightarrow BH}(xs,n,M_D)=F\,\frac{1}{M_{D}^2} \left[\frac{2^n
\pi^{n-1}\,\Gamma \left(\frac{n+3}{2}\right)} {(2+n)}\right]^{\frac{2}{n+1}}\,
\left(\frac{\sqrt{xs}}{M_{D}}\right)^{\frac{2}{n+1}}\,,
\label{partcross}
\end{equation}
where $F$ is a form factor.

It should be noted that $M_{BH,min}$ is not necessarily equal to the minimum
allowed mass of the BH, $\bar{M}$, and $M_{BH,min} \geq \bar{M} \geq M_D$ .
$M_{BH,min}$ and $\bar{M}$ depend on quantum gravity physics and cannot
presently be determined. $x_m$ is generally assumed to be a constant parameter
of order one. For spherically symmetric BHs a justification for this choice is
provided by the following semiclassical argument \cite{Anchordoqui:2003jr}: For
$M_{BH,min}/M_D\gtrsim 3$ and $n\geq 5$, the Hawking entropy of the BH is
larger than 10, and therefore strong gravitational effects can be neglected.
The semiclassical results based on Eq.~\refb{totcross} are then extrapolated
for $M_{BH,min}/M_D\lesssim 3$ with the assumption that the BH or its Planckian
progenitor decays on the brane, whatever the quantum theory of gravity may be.
As $\bar{M}$ and $M_D$ are not necessarily equal, caution is required in
extrapolating down the semiclassical results to $M_{BH,min}\sim \bar{M}\sim
M_D$, where the physics is unknown and quantum effects are important
\cite{Giudice:2001ce}. The entropy estimation is based on Hawking's
semiclassical theory and is not valid at energies a few times above the
fundamental scale. For example, it has been shown that the existence of a
minimum length could dramatically increase the value of $\bar{M}$, and thus
also increase $x_m$ \cite{Cavaglia:2003qk}. Finally, the estimate simply refers
to static spherically symmetric BHs, and may be drastically affected if the
geometry of the BH or its Planckian progenitor is different. 

\begin{figure}
\ybox{0.45}{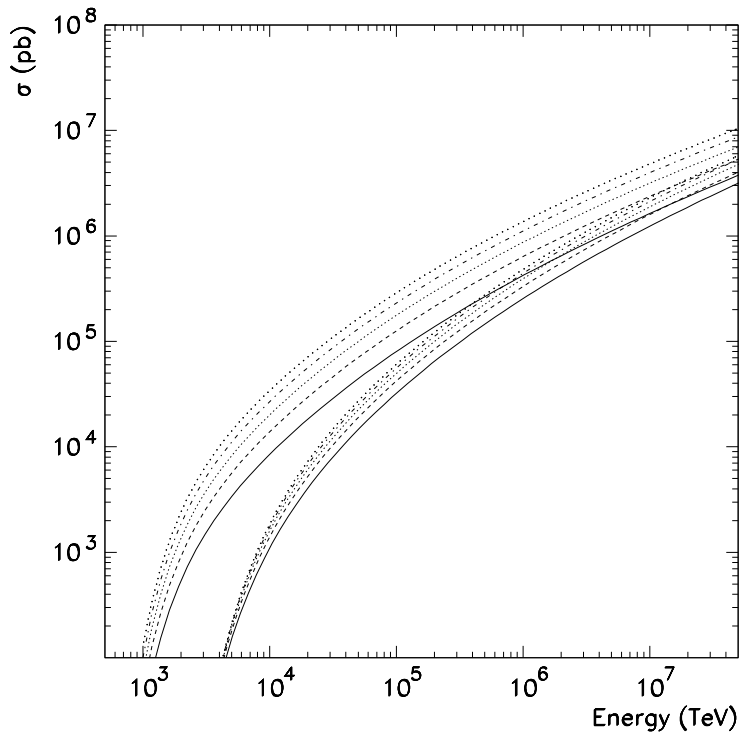}
\caption{BH cross section ($M_{BH,min} = M_D= 1.0$ TeV) for BD approximation
(left upper curves) and for YN formalism (right lower curves). For each group,
$n=3$ \ldots 7 from below. $Q = r_s^{-1}$ and and CTEQ6 PDFs is used.} 
\label{bdyn} 
\end{figure}

The form factor depends in principle on the energy of the process, on the
gravitational scale, on the geometry and number of extra dimensions, and on the
geometry and physical properties of the gravitational object. Some of the
physical parameters that can affect the form factor are angular momentum,
charge, geometry of the trapped surface, quantum corrections to classical
gravity, unknown effects of super-Planckian particle physics, structure and
topology of the compactified dimensions. With the lack of further insight, most
authors simply set $F=1$. Yoshino and Nambu (YN) \cite{Yoshino:2002tx}
numerically investigated the formation of the BH apparent horizon. In the YN
approach $F$ is a  numerical factor depending on the number of extra dimensions
ranging from $\approx 0.65$ ($n=0$) to $\approx 1.88$ ($n=7$). YN gives a
relation between the impact parameter and the mass of the BH which is formed in
the collision. The result is that the mass of the BH decreases as the impact
parameter increases up to a maximum value. This behavior affects the
computation of the total cross section by requiring the lower bound of the
integral in Eq.~\refb{totcross} to depend on the impact parameter. The total
cross section in the YN approach is \cite{Anchordoqui:2003jr}
\begin{equation}
\sigma'_{\nu p \to BH}(s,x_m,n,M_D) = \sum_{i}\int_0^1 2z dz\int_{x_m'}^{1} dx
\, q_i(x,-Q^2) \, \sigma_{\nu i \rightarrow BH}(xs,n,M_D) \,,
\label{totcrossYN}
\end{equation}
where $z$ is the impact parameter normalized to its maximum value and
$x_m'=x_m/y^2(z)$, $y(z)$ being the fraction of c.m.~energy that is trapped
into the BH. The YN approach lowers the BD cross section (Fig.~\ref{bdyn}).

The YN result still relies on a number of assumptions that may affect the final
estimate. (For recent criticisms, see Ref.~\cite{Vasilenko:2003ak}.) The
incoming partons, for example, are modeled as classical neutral point-like
particles. Partons carry color and EM charge, and it has been shown that the
physics of collisions between charged particles is quite different from that of
uncharged ones \cite{Casadio:2001wh}. Moreover, it is not clear what
constitutes the energy that is not trapped inside the horizon. Recent studies
seem to suggest that gravitational emission can account only for a part of the
missing BH energy \cite{Berti:2003si}. This could signal that the physics of
trans-Planckian collisions is more complex than the simple semiclassical
picture.

Another source of uncertainty in Eq.~(\ref{totcross}) and
Eq.~(\ref{totcrossYN}) comes from the ambiguity in the definition of the
momentum transfer for a BH event \cite{Emparan:2001kf}. The latter is usually
chosen to equal the BH mass or the inverse of the Schwarzschild radius.
However, there are no definite arguments to prefer either one or to exclude
alternative choices. The uncertainty due to the ambiguity in the definition of
the momentum transfer is evaluated as $\sim 10 - 20$\% 
\cite{Anchordoqui:2001cg}.

Finally, a minor but additional source of uncertainty in the total cross
section is due to the PDFs. Different sets of PDFs are defined in the
literature. The PDFs are not known for momentum transfer higher than a given
value (see, e.g.\ Ref.\ \cite{Pumplin:2002vw}), which is lower than the
momentum transfer expected in BH formation. Equation (\ref{totcross}) and
Eq.~(\ref{totcrossYN}) are calculated by imposing a cut-off at this energy.
They also suffer from uncertainties at any momentum transfer that can
contribute to the reduction in the total cross section. The minimum uncertainty
on the BH total cross section due to the PDFs can be estimated for a given
distribution. The CTEQ6 distribution gives an uncertainty of $\sim 3 - 4$\%, a
value that does not include the uncertainty due to the cutoff on the momentum
transfer nor the uncertainty introduced by the use of different sets of PDFs.
The MRST distribution \cite{Martin:2003tt} for $M_D=1.0$ TeV and
$M_{BH,min}/M_D=3$ gives results about $\sim 10 - 15$\% lower than the CTEQ
distribution.
\section{Effect of cross section uncertainties on $M_D$ bounds}
The BD approximation may be different than the actual cross section due to
these uncertainties. To give a concrete example of how the determination of the
cross section affects the $M_D$ bounds, let us look at a case where the parton
cross section is arbitrarily reduced by a constant factor of order one. The
cause of this reduction could be any one or a combination of the uncertainties
previously discussed.

\begin{figure}
\ybox{0.4}{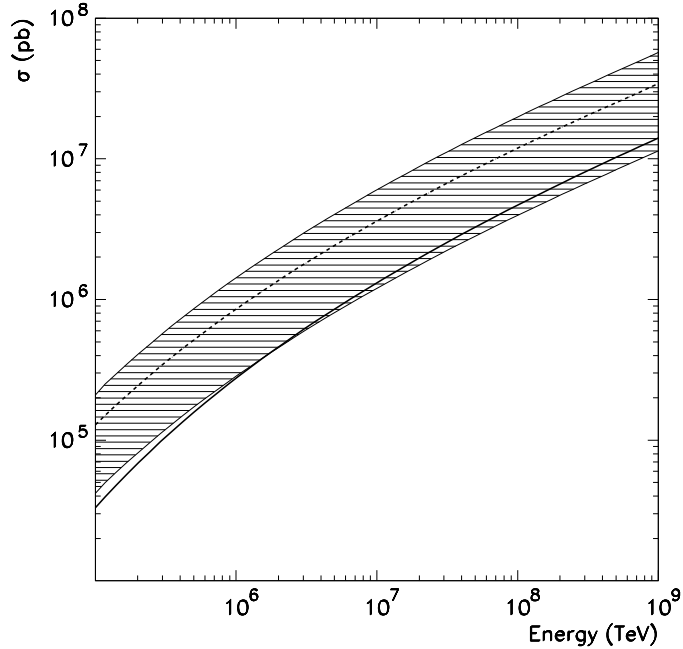}
\caption{The cross section for $M_D = 1.0$ TeV (thick solid line) and for $M_D
= 0.72$ TeV (dashed line). Shaded band denotes $F=1\pm 2/3$ for $M_D = 0.72$
TeV. The cross section of $M_D = 1$ TeV with $F=1$ lies within this band. CTEQ6
is used.}
\label{uncertain}
\end{figure}

Consider high energy primary neutrinos with energy $E_\nu=10^6 - 10^8$ TeV,
$M_{\rm BH,min}=3M_D$ and $n=5$. For the sake of simplicity, we temporarily
neglect the YN results and consider Eq.~\refb{totcross}. This is sufficient for
a rough estimation. The total cross section \refb{totcross} for $M_D=1$ TeV and
$F=1$ lies within the shaded band of the total cross section for $M_D=0.72$ TeV
and $F=1\pm 2/3$. This is shown graphically in Fig.~\ref{uncertain}. Therefore,
what is interpreted as the $M_D=1.0$ TeV bound in the naive BD approximation
can actually be a less stringent bound if the parton cross section is
two-thirds smaller: $M_D = 0.72$ TeV, the lower bound from collider experiments
derived from virtual graviton exchange \cite{collider, Giudice:2003tu}. The
situation is made even more complicated by the presence of the unknown
parameter $M_{BH,min}$. Increasing $M_{BH,min}$ corresponds to decreasing the
BH cross section. The lower bound on the eleven-dimensional $M_D$ for
$M_{BH,min}/M_D=5$ is approximately half the bound for $M_{BH,min}/M_D=1$. 

\section{Number of events and neutrino flux}
The uncertainties in the BH cross section listed above apply to BH formation by
UHECR events as well as by particle colliders. In cosmic ray events, however,
the unknown neutrino flux adds further uncertainties. In the YN approach, the
number of neutrino-nucleon BH events detected by a cosmic ray detector in time
$T$ is \cite{Anchordoqui:2003jr}:
\begin{equation}
N =N_A T \sum_{i} \int dE_{\nu}\, \int_0^1 2z dz\int_{x_m'}^{1} dx \,
q_i(x,-Q^2) \, \sigma_{\nu i \rightarrow BH}(xs,n,M_D) \,
\frac{d\Phi}{dE_{\nu}} \, A(y E_{\nu})\,,
\label{events}
\end{equation}
where $A(yE_{\nu})$ is the experiment acceptance for an air shower energy
$yE_\nu$, $N_A$ is Avogadro's number, and $d\Phi/dE_{\nu}$ is the source flux
of neutrinos. 

The cosmogenic neutrino flux is considered to be the most reliable source of
neutrinos. In this model \cite{berezinsky1} neutrinos are produced from
ultrahigh energy protons interacting with the ubiquitous cosmic microwave
background. However, this is not fully guaranteed as the existence of the
cosmogenic neutrino flux relies on the assumption that cosmic rays with
energies above $10^8$ TeV are extragalactic protons. Neither the source nor
composition of cosmic rays above $10^8$ TeV are known (see, e.g.,
Ref.~\cite{Ave:2000nd, Olinto:2000sa}). If these are heavy nuclei or photons,
or Lorentz invariance is violated \cite{lorentzv}, there may be no cosmogenic
neutrino flux at ultrahigh energies. Even if UHECRs are protons from
extragalactic sources, cosmological evolution, spatial distribution, abundance,
and injection spectrum of UHECR sources can change the cosmogenic neutrino flux
by an order of magnitude. In order to derive a conservative lower bound on
$M_D$, the lower end of plausible cosmogenic neutrino fluxes should be used in
Eq.~\refb{events}. 

\begin{figure}
\ybox{0.45}{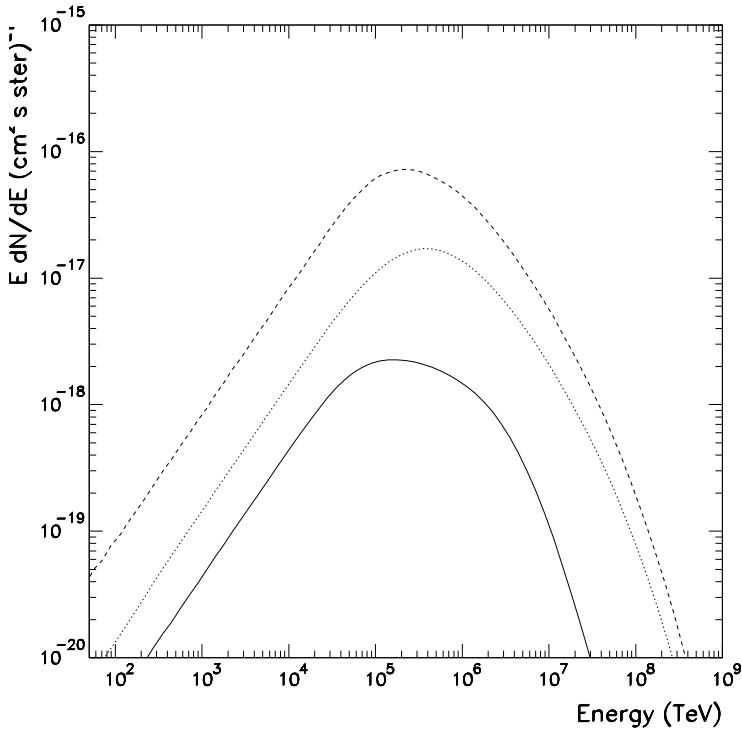}
\caption{Cosmogenic neutrino fluxes for the $\nu_{\mu}$ family. PJ (dashed
line), ESS (dotted line), and our model (solid line) fluxes from the top.}
\label{flux}
\end{figure}

Here we consider a model that gives a relatively low neutrino flux in agreement
with observations \footnote{We thank Todor Stanev for providing the data.}. The
flux is calculated following the procedure of Ref.~\cite{Engel:2001hd}. The
source spectrum is proportional to $E^{-2.6} \times \mathrm{exp} \it (-E/E_c)$,
where $E_c = 10^{8.5}$ TeV is the cutoff energy, and normalized to the cosmic
ray luminosity at $10^7$ TeV. There is no cosmological evolution and the
redshift integration is from $z = 0.05 - 8.00$. The parameters used are
consistent with observations: A spectral index of 2.6 is the best fit to the
highest energy data \cite{DeMarco:2003ig} and the cutoff energy is limited by
the highest energy cosmic ray event observed to date. Since the ultrahigh
energy proton sources are unknown, their evolution cannot be determined. No
evolution is a reasonable assumption for the lower end of the neutrino flux
that should be considered. Our model flux is in good agreement with the lower
bound obtained from cosmic ray data analysis \cite{Fodor:2003ph}. Figure
\ref{flux} compares this flux to two other fluxes, by Protheroe and Johnson
(PJ) \cite{Protheroe:1995ft} and Engel, Seckel, and Stanev (ESS)
\cite{Engel:2001hd}. Our flux is lower by at least an order of magnitude than
the PJ flux on the most relevant range of neutrino energies.

The sensitivity of the detector for horizontal air showers also affects the
computation of the lower bounds on $M_D$. For example, the Pierre Auger
Observatory \cite{Auger} uses 1.2 m high water Cherenkov detectors whereas
AGASA \cite{AGASA} uses 5 cm thick scintillators. From purely geometrical
arguements, AGASA's detection capability rapidly goes down at large zenith
angle. On the other hand, AGASA has a lower detector trigger threshold which
gives higher sensitivity at lower energies. Scaling the aperture by the
experiment size is an approximation that should be further improved.
\section{Realistic bounds on $M_D$}

The effect of the cosmogenic neutrino flux on $M_D$ lower bound is best
illustrated by comparing the conservative flux of Sect.~IV and the PJ flux. We
performed a systematic analysis of the number of events using the above
cosmogenic neutrino fluxes for different experiments (AGASA \cite{Yoshida},
HiRes \cite{Abu-Zayyad:1999xa}, RICE \cite{Kravchenko:2001id}) and different
choices of parameters in the total cross sections from Eq.~\refb{totcross} and
\refb{totcrossYN}. For simplicity we used the apertures given by
Refs.~\cite{Anchordoqui:2001cg, Anchordoqui:2002vb, Anchordoqui:2003jr}. Though
RICE is an experiment looking into ``ice showers" rather than EASs, the
technique used in EAS experiments is applicable. The use of the PJ flux gives a
lower bound of $M_D \geq 1.0 - 1.4$ TeV, which is comparable to and sometimes
more stringent than collider bounds. The most optimistic case with our model
flux is when $Q = M_{BH}$ and $M_{BH,min}= M_D$. The lower bounds on the
fundamental scale range from $M_D=0.3$ ($M_{BH,min}=M_D$) to $M_D <0.2$ TeV
($M_{BH,min}=5M_D$). These limits are much lower than the collider limits
\footnote{The computations were performed using the CTEQ6 PDFs and YN
approximation. The use of MRST PDFs give results identical within an
uncertainty of 0.05 TeV and the BD approximation increases $M_D$ by $0.05 -
0.15$ TeV.} \cite{collider, Giudice:2003tu}. 

We also considered changing the cosmological evolution of the conservative flux
to be that of Ref.~\cite{Waxman:1995dg}, i.e.
\begin{equation}
\begin{array}{ll}
(1 + z)^3  \hspace{6cm}(z<1.9)\\
(1 + 1.9)^3 \hspace{5.7cm}(1.9<z<2.7)\\
(1 + 1.9)^3 \rm exp[(2.7-z)/2.7] \, \hspace{2.5cm} (2.7<z) \hspace{1cm}.
\end{array}
\label{gzkeqn}
\end{equation}
This increases the flux by about an order of magnitude. It is lower than the
ESS flux for $E_\nu \gtrsim 10^6$ TeV, and is considerably less by a couple of
factors than the PJ flux in all energy ranges. The lower bounds increase by
about 0.4 TeV, but are still below the collider bounds.
\section{Conclusions}
We have reviewed the method of constraining the fundamental Planck scale from
nonobservation of EASs and discussed the uncertainties on the $M_D$ lower bound
coming from the BH cross section and neutrino flux. The major source of
uncertainties in the cross section is due to the lack of a definite theory of
trans-Planckian scattering. Trans-Planckian gravity is not known at present to
formulate a reliable model for BH formation. The BD approximation and its
variants seem to provide a reasonable model for the BH formation cross section
at parton level. However, the naive BD cross section relies on a number of
crude assumptions and suffers from uncertainties that cannot be estimated.
Neglected quantum gravity effects are expected to be relevant. 

A conservative estimation of $M_D$ lower bounds is obtained by taking into
account these caveats and considering the range of cosmogenic neutrino fluxes
which are compatible with observations. The conclusion is that in the absence
of independent determination of the cosmogenic neutrino flux and of reliable
theoretical computation of the BH cross section, nonobservation of BH-induced
deeply penetrating EASs gives lower bounds on the fundamental gravitational
scale less stringent than present collider bounds.

\section*{Acknowledgements}

We thank T.~Stanev for helpful discussions and providing the flux data, and
L.~Anchordoqui, M.~Ave, A.~Chou, U.~D'Alesio, H.~Goldberg, S.~C.~Park,
T.~Weiler, T.~Yamamoto, H.~Yoshino, and R.~P.~Mashimaro for helpful discussion.
M.~C. thanks M.~Anselmino and the Department of Theoretical Physics of the
University of Torino for kind hospitality. This work was supported in part by
the NSF through grant AST-0071235 and DOE grant DE-FG0291-ER40606 at the
University of Chicago and at the Kavli Institute for Cosmological Physics by
grant NSF PHY-0114422. KICP is a NSF Physics Frontier Center.

\section*{Note added in Proof}
                                                                                
Recent work by Hossenfelder \cite{Hossenfelder:2004} calculates the BH cross
section suppression from minimal length at LHC. It is found that the ratio
between the total cross section with and without minimal length effects is
approximately 0.19 for the expected LHC energy, and increasing at higher
energies. This result strenghtens our arguments of section II, namely that the
presence of a minimum length may substantially reduce the BD cross section.

In addition, progress in neutrino flux estimates with cosmic ray primaries
other than protons have shown how uncertain the neutrino flux can be (see
hep-ph/0409316 and hep-ph/0407618).

\end{document}